\documentclass[preprintnumbers,article,amsmath,amssymb,floatfix,10pt,prd,onecolumn,
superscriptaddress,nofootinbib]{revtex4}
\usepackage{bbm}
\usepackage{amsfonts}
\usepackage{mathrsfs}
\usepackage{latexsym}

\usepackage{epsfig}
\usepackage{graphicx}
\usepackage{epstopdf}
\usepackage{placeins}
\usepackage{float}

\usepackage{amssymb}
\usepackage{amsmath}

\usepackage{dcolumn}
\usepackage{bm}
\usepackage{color}
\usepackage{comment}
\usepackage{xcolor}

\begin{document}
\title{\bf Study of Embedded Class-I Fluid Spheres in $f(R,T)$ Gravity with \\ Karmarkar Condition}
\author{Zoya Asghar}
\email{zoyaasghar16411@gmail.com}\affiliation{National University of Computer and
Emerging Sciences,\\ Lahore Campus, Pakistan.}
\author{M. Farasat Shamir}
\email{farasat.shamir@nu.edu.pk; farasat.shamir@gmail.com}\affiliation{National University of Computer and
Emerging Sciences,\\ Lahore Campus, Pakistan.}
\author{Ammara Usman}
\email{ammarausman88@ymail.com}\affiliation{Lahore Leads University, Lahore, Pakistan.}
\author{Adnan Malik}
\email{adnan.malik@zjnu.edu.cn; adnan.malik@skt.umt.edu.pk }
\affiliation{School of Mathematical Sciences, Zhejiang Normal University, \\Jinhua, Zhejiang, China.}
\affiliation{Department of Mathematics, University of Management and Technology,\\ Sialkot Campus, Lahore, Pakistan}

\begin{abstract}
In this article, we explore some emerging properties of the stellar objects in the frame of the $f(R,T)$ gravity by employing the well-known Karmarkar condition, where $R$ and $T$ represent Ricci scalar and trace of energy momentum tensor respectively. It is worthy to highlight here that we assume the exponential type model of $f(R,T)$ theory of gravity $f(R,T)=R+\alpha(e^{-\beta R}-1)+\gamma T$ along with the matter Lagrangian $\mathcal{L}_{m}=-\frac{1}{3}(p_{r}+2 p_{t})$ to classify the complete set of modified field equations. We demonstrate the embedded class-I technique by using the static spherically symmetric line element along with anisotropic fluid matter distribution. Further, to achieve our goal, we consider a specific expression of metric potential $g_{rr}$, already presented in literature, and proceed by using the Karmarkar condition to obtain the second metric potential. In particular, we use four different compact stars, namely $LMC~X-4,$ $EXO~1785-248,$ $Cen~X-3$ and $4U~1820-30$ and compute the corresponding values of the unknown parameters appearing in metric potentials. Moreover, we conduct various physical evolutions such as graphical nature of energy density and pressure progression, energy constraints, mass function, adiabatic index, stability and equilibrium conditions to ensure the viability and consistency of our proposed model. Our analysis indicates that the obtained anisotropic outcomes are physically acceptable with the finest degree of accuracy.
\end{abstract}
\maketitle
\section{Introduction}
The study of relativistic stellar objects such as white dwarfs, black holes and neutron stars has gained most appreciation among cosmologists within the last few decades. These stars originate as a result of gravitation collapse of massive objects. 
All these kinds of stars are popularly known as degenerate stars, other than black holes. Neutron stars are stellar structures that originated when a giant star exhausts its fuel and collapses. These stars are too dense because of their tiny size and immensely massive structure. The mass of neutron stars lies between $1.1M_{\odot}$ and $2.01M_{\odot}$. Whereas, when a star, such as the Sun, runs out of nuclear fuel, it becomes a white dwarf. The white dwarf's solar mass is generally within $0.17M_{\odot}$ and $1.33M_{\odot}.$ On the other hand, its density varies between $104$ to $107 g/cm^3$. The white dwarf will change into a black dwarf when it has expelled all of its heat. Also, the most captivating and strangest objects in space are black holes. They are extraordinarily dense, with such a powerful gravitational pull that not even light can pass through them. When a massive star (approximately three times the size of the sun) falls, it forms a black hole. Einstein field equations $(EFE)$ are considered as the most fruitful approach in exploring the complicated nature of compact objects. In this regard, the very first exact solution of $EFE$ was proposed by Schwarzschild \cite{Sch}. Later on, a detailed analysis of stellar structures was presented by Tolman \cite{Tol} and Oppenheimer \cite{opp1}. They declared that the physical properties of compact stars reflected the connection between gravitational force and internal pressure, which maintain a state of equilibrium. Further, the internal core of stellar structure was discussed by Baade and Zwicky \cite{Baad}. In the beginning, cosmologists believed that the inner structure of compact objects was only narrated by perfect fluid. Although isotropy may contain suitable attributes, it still does not present the usual characteristics of compact stars. In cosmology, the anisotropic fluid gains much attention as an alternative of isotropic matter distribution. Anisotropic matter used to describe the phase development and internal characterization of the stellar configurations. Anisotropy is considered in the fluid due to the amalgamation of different categories of fluids, magnetic field, viscosity, rotation etc. Bowers and Liang \cite{Bow} first proposed the theory of non-zero anisotropy in the star structure. Further, the concept of high density of the stellar structure demonstrates the nature of anisotropy at the center of the stellar object was initially introduced by Ruderman \cite{Rud}. The anisotropy in stellar structure might be revealed due to presence of phase transition, a star having pion condensation \cite{Saw}, electromagnetic field \cite{Put,Rei,Mar}, survival of solid core etc.
Generally, in case of anisotropic distribution the pressures are partied into radial and transverse expressions.

In astrophysics, the mysterious expansion of our universe is one of the most eye-catching topic and it is a common belief among cosmologists that this expansion relies on the dark energy $(DE)$ and dark matter $(DM)$ which conserve the negative pressure. Although general relativity $(GR)$ is suitable in explaining the complicated nature of compact stars, however, in the case of $DM$ and $DE$, $GR$ failed to deliver the realistic outcomes. That's why in order to examine the mysterious nature of $DE$, we need modified gravitational theories. The popular modified gravitational theories are listed as $f(R)$, $f(R,T)$, $f(T)$, $f(G)$, $f(R,G)$ and $f(R, \phi, X)$ theories of gravity \cite{Sham1,Sham2,BLi2,Har,Cap2,Lov,Har2,Ikr,ad5,ad6,ad7,ad9,Zoy}. Buchdahl explored the $f(R)$ gravity \cite{Buch}, which is recognized as the best fundamental and eminent modified theory. Recently, Harko \cite{Har} introduced a captivating modification of $f(R)$ theory of gravity, well-known as $f(R,T)$, which is basically a amalgamation of trace of stress-energy tensor $(SET)$ and Ricci curvature. Later on, Adhav \cite{Adh} investigated the  precise solutions of $EFE$ for the locally rotationally symmetric Bianchi Type-I line element in the background of $f(R,T)$ theory of gravity. Further, Ahmed and Paradhan \cite{Nai} studied the Bianchi Type-V cosmology by using cosmological constant for the $f(R,T)$ gravity. Moreover, the evaluation of deformation and consistency criteria of $f(R,T)$ gravity along with thermodynamic laws was discussed by Sharif and Zubair \cite{Sha1,Zub}. In this regard, Pretel et al. \cite{Pre} explored the behavior of the stellar system in the $f(R,T)$ model background.
Furthermore, Waheed et al. \cite{WGZA} discussed the viable embedded class-I solutions of stellar objects in the light of $f(R,T)$ gravity by applying Karmarkar condition and their attained outcomes are quite realistic and physically acceptable.

For a well-behaved stellar structure, cosmologists applied an analytical technique of $EFE$ and considered a four-dimensional manifold family which converts into an Euclidean space. Embedded problem of the compact star was first studied by Schlai \cite{Schl} in 1871. Later the very first theorem in regard to isometric embedded was discovered by Nash \cite{Nas} and the modification of this theorem was later presented by Gunther \cite{Gunt}.
Gupta and Sharma \cite{Gup} considered the plane metric to get the solution of embedded class. The embedded classes generate a differential equation in spherical symmetric spacetime which combines both the metric potential components, $g_{rr}$ and $g_{tt}$, and is called the Karmarkar condition. The mentioned condition was introduced by Karmarkar \cite{KR} and is defined as $R_{1414}R_{2323}=R_{1212}R_{3434}+R_{1224}R_{1334}$.
Naz et al. \cite{Naz} investigated the embedded class-I outcomes of stellar system for $f(R)$ theory.
In this paper, we aim to expand the idea of Naz et al. \cite{Naz} in $f(R,T)$ theory context and check the physical viability of our system by considering four different compact star namely, $LMC~X-4,$ $EXO~1785-248,$ $Cen~X-3$ and $4U~1820-30$. To gain our goal, we assume the exponential type $f(R,T)$ theory model, i.e. $f(R,T)=R+\alpha(e^{-\beta R}-1)+\lambda T$ \cite{Cog,Har} along with the matter Lagrangian $\mathcal{L}_{m}=-\frac{1}{3}(p_{r}+2 p_{t})$ \cite{Har}. Moreover, we take a specific expression for the metric potential $g_{rr}$ \cite{BSM}, already presented in literature, and determine the second metric potential by utilizing the famous Karmarkar condition. The layout of our profile is as follows: In the coming section, the modified field equations along with a realistic model of $f(R,T)$ theory have been discussed in the light of Karmarkar condition. The unknown constants have been calculated in the background of matching conditions in portion III. We analyze the graphical nature of the relativistic objects in the IV sector. In the last section, we present the final consequence of our work.
\section{Modified $f(R,T)$ Field Equations}
Firstly, we discuss all the important assumptions taken for our work, in order to develop the necessary configuration of the framework of modified $f(R,T)$ theory. The generalized action of $f(R,T)$ theory is describe as
\begin{equation}\label{0}
S= \int\Big(\frac{1}{2\kappa}f(R,T)+\mathcal{L}_{m}\Big)\sqrt{-g}d^4x,
\end{equation}
here, $R$ symbolizes Ricci scalar and $T$ represents trace of $SET$, $\mathcal{L}_{m}$ is Lagrangian matter, $\kappa=\frac{8\pi G}{c^{4}}$ shows coupling constant and $g$ exhibits the determinant of metric $g_{\eta\zeta}$.
The modified field equation of $f(R,T)$ theory is defined as
\begin{equation}\label{1}
f_{R}(R,T) R_{\eta\zeta}-\frac{1}{2} f(R,T)g_{\eta\zeta}+(g_{\eta\zeta} \Box-\nabla_\eta\nabla_\zeta)f_R(R,T)= 8\pi T_{\eta\zeta}-f_T(R,T)T_{\eta\zeta}-f_T (R,T) \Theta_{\eta\zeta},
\end{equation}
where, $f_R(R,T)=\frac{df(R,T)}{dR}$ and $f_T(R,T)=\frac{df(R,T)}{dT}$, whereas, $\Box\equiv\nabla_{\eta}\nabla^{\zeta}$ indicates the D'Alembertian symbol and $\nabla_\eta$ denotes covariant derivative and $g_{\eta\zeta}$ is metric tensor.
The covariant derivative Eq. (\ref{1}) \cite{Barri} shows
\begin{equation}\label{2}
\nabla^\eta T_{\eta\zeta} = \frac{f_{T}(R,T)}{8\pi-f_{T}(R,T)}\Big((T_{\eta\zeta}+\Theta_{\eta\zeta})\nabla^\eta \ln f_{T}(R,T)+\nabla^\eta \Theta_{\eta\zeta}-\frac{1}{2}g_{\eta\zeta}\nabla^\eta \mathcal{T}\Big),
\end{equation}
which demonstrates that the $SET$ in $f(R,T)$ does not confirm the conservation law as in $GR$. The field equations (\ref{1}) can be defined in a $GR$ form as
\begin{equation}\label{4}
G_{\eta\zeta}=\frac{1}{f_{R}(R,T)}\Big[8\pi T_{\eta\zeta}+\frac{1}{2}(f(R,T)-R f_{R}(R,T))g_{\eta\zeta}-f_{T}(R,T)(T_{\eta\zeta}+\Theta_{\eta\zeta})-(g_{\eta\zeta}\Box- \nabla_\eta\nabla_\zeta)f_{R}.(R,T)\big].
\end{equation}
The term $\Theta_{\eta\zeta}$ is called scalar expansion which is represent as $\Theta_{\eta\zeta}=-2 T_{\eta\zeta}-P g_{\eta\zeta},$ whereas, $P$ is considered as $P=\frac{1}{3}(p_{r}+2 p_{t})$ for an anisotropic pressure \cite{Har}. Further, we consider a static and spherical spacetime line element to represent the interior core of the compact star i.e.
\begin{equation}\label{5}
ds^{2}=e^{\nu(r)}{dt}^{2}-e^{\lambda(r)}{dr}^{2}-r^{2}({d\theta}^{2}+sin^{2}\theta{d\phi}^{2}).
\end{equation}
Here, ${\nu(r)}$ and ${\lambda(r)}$ are functions of radial coordinates only. Moreover, $T_{\eta\zeta}$ symbolizes the $SET$ and, in case of anisotropic sphere, $T_{\eta\zeta}$ is given by
\begin{equation}\label{3}
T_{\eta \zeta}=(\rho+p_{t})u_{\eta}u_{ \zeta}-p_{t}g_{\eta \zeta}+(p_{r}-p_{t})\mathcal{X}_{\eta}\mathcal{X}_{ \zeta}.
\end{equation}
In the above equation, the terms $\rho$, $p_{r}$ and $p_{t}$ represent energy density and pressure components (radial and transverse) respectively. Here, $u_{\eta}$ and $\mathcal{X}_{\eta}$ denote four velocity vectors, which can be defined by the relation i.e. $u^{\eta}u_{\eta}=-\mathcal{X}^{\eta}\mathcal{X}_{\eta}=1$. It is important to mention here that Eq. (\ref{3}) corresponds to an anisotropic perfect fluid, while Eq. (\ref{2}) corresponds to a break of usual conservation laws. Thus the usage of these two equations together is prohibited as the conservation law for stress-energy tensor is not obeyed, i. e. the covariant derivative of stress-energy tensor does not seems to vanish. However, one can put some constraints to Eq. (\ref{2}) to obtain standard conservation equation for stress-energy. This has been taken into account by taking the extra force $\mathcal{F}_{frt}=-\frac{2\gamma}{3(8\pi-\gamma)}\frac{d}{dr}(3\rho-p_{r}-2 p_{t})$ (due to the right hand side of Eq. (\ref{2})) to be equal to zero.

To probe the stability and consistency of the stellar system, we consider an exponential model of $f(R,T)$ theory \cite{Cog,Har} which helps us in developing a comprehensive understanding about the physical structure of the stellar star.The considered $f(R,T)$ theory model is defined as
\begin{equation}\label{6}
f(R,T)=R+\alpha(e^{-\beta R}-1)+\gamma T,
\end{equation}
where, $\alpha,$ $\beta$ and $\gamma$ are any arbitrary constants. Furthermore, the spacetime metric tensors shows the family of an embedded class-I, if its fulfil the famous Karmarkar condition \cite{KR}, which is given as
\begin{equation}\label{7}
 R_{1414}R_{2323}=R_{1212}R_{3434}+R_{1224}R_{1334},
\end{equation}
with $R_{2323}\neq 0$.
By putting the required Riemann tensor values in Eq. (\ref{7}), we receive the succeeding differential form
\begin{equation}\label{8}
 \lambda^{'}\nu^{'}-2\nu^{''}-{\nu^{'}}^{2}=\frac{\lambda^{'}\nu^{'}}{1-e^{\lambda}},
\end{equation}
such that $e^{\lambda}\neq1$. Integrating the above differential Eq. (\ref{8}), we get
\begin{equation}\label{9}
  e^{\nu}=\Big[{{\big(A+B\int{\sqrt{e^{\lambda}-1} dr}\big)}^{2}}\Big],
\end{equation}
here, $A$ and $B$ are integrating constants. To determine the solution of embedded class-I, we assume particular value of the metric potential $g_{rr}=e^{\lambda(r)}$ \cite{BSM} i.e.
\begin{equation}\label{10}
  e^{\lambda}=1+ \frac{a^{2}r^{2}}{(1+br^{2})^{4}},
\end{equation}
where, $a$ and $b$ are any non-zero constant parameters. Substituting Eq. (\ref{10}) in Eq. (\ref{9}), metric potential $e^{\nu}$ takes the form
\begin{equation}\label{11}
  e^{\nu}=\Big[{\Big(A-\frac{aB}{2b(1+br^{2})}\Big)}^{2}\Big].
\end{equation}
\begin{figure}[h!]
\begin{tabular}{cccc}
\epsfig{file=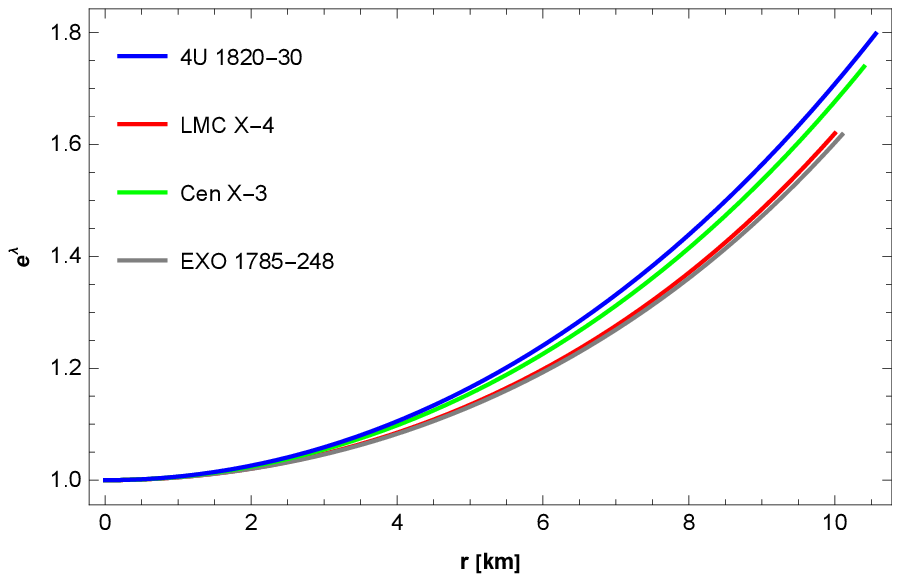,width=0.38\linewidth} &
\epsfig{file=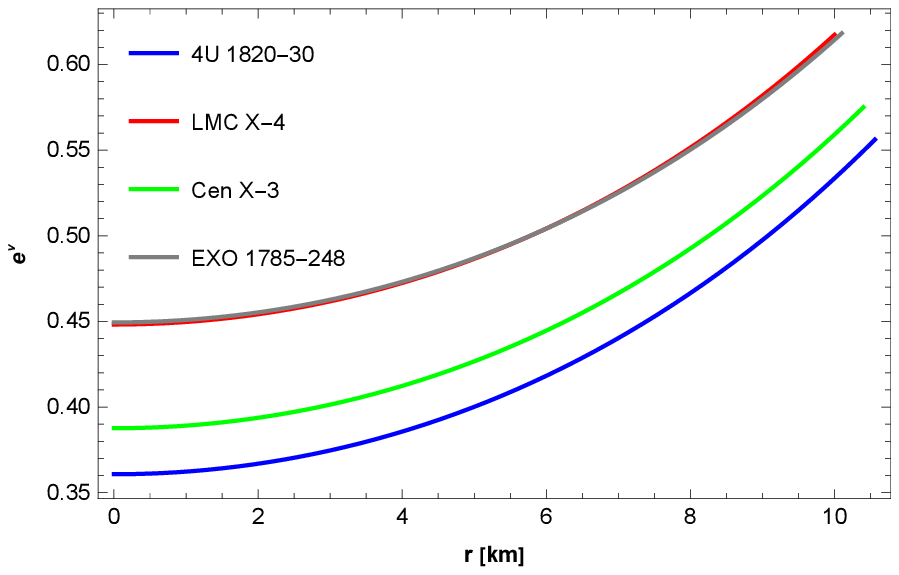,width=0.38\linewidth} &
\end{tabular}
\caption{{Behaviour of metric potential $g_{rr}$ and $g_{tt}$}.}
\label{Fig:1}
\end{figure}
\FloatBarrier
Moreover, to explore the effects of our chosen $f(R,T)$ model, we studied the  mandatory requirements for the metric tensors of static spherically symmetric spacetime, that is $e^{\lambda(0)}=1$ and $((e^{\lambda(r)})')_{r=0}=0$. For a consistent $f(R,T)$ gravity model, the graphical behavior of $g_{rr}$ should be free of singularity, regular and monotonically increasing. It is notable from Fig. \ref{Fig:1} that the physical behavior of $e^{\lambda(r)}$, which is monotonically increasing and reaches the maximum value at the boundary. Thus, by utilizing Eqs. (\ref{6}), (\ref{10}) and (\ref{11}) the modified $f(R,T)$ field equation given in Eq. (\ref{4}) yields
\begin{eqnarray*}
\nonumber &&
\rho= \frac{1}{24 (\gamma +4 \pi )} \Big(\frac{1}{(\gamma +8 \pi ) r \left(\frac{a^2 r^2}{\left(b r^2+1\right)^4}+1\right)}\big(h_{1} \big(\frac{2 a^2 r \left(3 b r^2-1\right) \left(\gamma  r \nu '-16 (\gamma +6 \pi )\right)}{\left(b r^2+1\right) \left(a^2 r^2+\left(b r^2+1\right)^4\right)}+\frac{16 a^2 (\gamma +6 \pi ) r}{\left(b r^2+1\right)^4}+\gamma  \nu ' \big(r \nu '~~~
\\&& +4\big)+2 \gamma  r \nu '')\big)- h_{2} \left(-\frac{2 a^2 (5 \gamma +48 \pi ) r^2 \left(3 b r^2-1\right)}{\left(b r^2+1\right) \left(a^2 r^2+\left(b r^2+1\right)^4\right)}+28 \gamma -3 \gamma  r \nu '+192 \pi \right)-2 (7 \gamma +48 \pi ) h_{3} r\big)+\alpha  \left(6-6 e^{\beta  -h_{4}}\right)~~~
\\&& -6 \alpha  \beta  h_{4} e^{\beta  -h_{4}}\Big),
\end{eqnarray*}
\begin{eqnarray*}
\nonumber &&
p_r= \frac{1}{24 (\gamma +4 \pi )}\Big(\frac{1}{(\gamma +8 \pi ) r \left(\frac{a^2 r^2}{\left(b r^2+1\right)^4}+1\right)}\Big(h_{1} \big(-\frac{2 a^2 \gamma  r \left(3 b r^2-1\right) \left(r \nu '(r)+8\right)}{\left(b r^2+1\right) \left(a^2 r^2+\left(b r^2+1\right)^4\right)}+\nu' \left(20 \gamma +\gamma  (-r) \nu '+96 \pi \right)-2 \gamma~~~
\\&&  r \nu ''\big)-\frac{16 a^2 (\gamma +6 \pi ) h_{1} r}{\left(b r^2+1\right)^4}+h_{2} \left(-\frac{2 a^2 (17 \gamma +96 \pi ) r^2 \left(3 b r^2-1\right)}{\left(b r^2+1\right) \left(a^2 r^2+\left(b r^2+1\right)^4\right)}+4 (7 \gamma +48 \pi )+3 (3 \gamma +16 \pi ) r \nu '\right)-10 \gamma  h_{3} r\Big)+~~~
\\&& 6 \alpha  \left(e^{\beta  -h_{4}}-1\right)+6 \alpha  \beta  h_{4} e^{\beta -h_{4})}\Big),
\end{eqnarray*}
\begin{eqnarray*}
\nonumber &&
p_t= \frac{1}{24 (\gamma +4 \pi )}\Big(\frac{1}{(\gamma +8 \pi ) r \left(\frac{a^2 r^2}{\left(b r^2+1\right)^4}+1\right)}\Big(h_{1} \big(\frac{2 a^2 r \left(3 b r^2-1\right) \left(4 (\gamma +12 \pi )+(5 \gamma +24 \pi ) r \nu '\right)}{\left(b r^2+1\right) \left(a^2 r^2+\left(b r^2+1\right)^4\right)}+\frac{8 a^2 \gamma  r}{\left(b r^2+1\right)^4}+\nu ' \big(8 (\gamma ~~~
\\&& +6 \pi )+(5 \gamma +24 \pi ) r \nu '\big)+2 (5 \gamma +24 \pi ) r \nu ''\big)+h_{2} \left(-\frac{2 a^2 (5 \gamma +48 \pi ) r^2 \left(3 b r^2-1\right)}{\left(b r^2+1\right) \left(a^2 r^2+\left(b r^2+1\right)^4\right)}+4 (\gamma +24 \pi )+3 (3 \gamma +16 \pi ) r \nu '\right)-~~~
\\&& 6 \alpha  (\gamma +8 \pi ) r \left(\frac{a^2 r^2}{\left(b r^2+1\right)^4}+1\right)+2 (7 \gamma +48 \pi ) h_{3} r\Big)+6 \alpha  e^{\beta -h_{4}} (\beta  h_{4} +1)\Big).
\end{eqnarray*}
Here,\\
\begin{eqnarray*}
 ~~~~~h_{1}=1-\alpha\beta e^{-\beta h_{4}},~~~~~~~~~~h_{2}&=&\frac{\partial h_{1}}{\partial r},~~~~~~~~~~h_{3}=\frac{\partial h_{2}^{2}}{\partial r^{2}}.
\end{eqnarray*}
$~~~~~~~~~~~~~~~~~h_{4}=\frac{2 a \left(-2 a^3 A b r^2 \left(b r^2+1\right)-a^2 B \left(b r^2-3\right) \left(b r^2+1\right)^3+a^4 B r^2+2 a A b \left(5 b r^2-3\right) \left(b r^2+1\right)^4-2 b B \left(b r^2-3\right) \left(b r^2+1\right)^6\right)}{\left(a^2 r^2+\left(b r^2+1\right)^4\right)^2 \left(2 A b \left(b r^2+1\right)-a B\right)},$

\section{Matching Conditions}
The interior boundary metric irrespective of intrinsic and extrinsic geometry of the stellar objects shall stay consistent. This statement validates that metric components are independent from coordinate systems over the boundary. In this regard, the Schwarzschild exterior geometry is always assumed to be the best choice of the junction constraints for stellar stars in the light of $GR$. In the background of spherical symmetric spacetime, Jebsen-Birkhoff’s theorem states that the field equations solutions must be static and asymptotically flat. When we discuss the improvised form of the $TOV$ equation \cite{Tol,opp1} having zero pressure and energy density, the extrinsic geometry solution may vary from the Schwarzschild’s solution in modified gravitational theories. On the contrary, this issue can be resolved in modified $f(R,T)$ theory by considering an appropriate model of the $f(R,T)$ theory. This mechanism violates Birkhoff's theorem in modified gravitational theories \cite{Fara}. A lot of investigation has already been done related to junction constraints which produces a noteworthy proposal related to Schwarzschild’s solution \cite{AVC,Coon,Gang,Mome}. Furthermore, the spherically symmetric spacetime solutions matches with the Schwarzschild’s exterior line element, defined as
\begin{equation}\label{12}
ds^{2}=\big(1-\frac{2M}{r}\big){dt}^{2}-\big(1-\frac{2M}{r}\big)^{-1}{dr}^{2}-r^{2}({d\theta}^{2}+sin^{2}\theta{d\phi}^{2}),
\end{equation}
here, $M$ denotes the mass of stellar star. At the boundary $r = R$, the continuity conditions of Eq. (\ref{5}) takes the succeeding form
\begin{equation}\label{13}
{g_{rr}}^{+}={g_{rr}}^{-},~~~~~~~~~~{g_{tt}}^{+}={g_{tt}}^{-},~~~~~~~~~~\frac{\partial{g_{tt}}^{+}}{\partial r}=\frac{\partial{g_{tt}}^{-}}{\partial r}~.
\end{equation}
Where, $(+)$ and $(-)$ demonstrate the exterior and interior geometry, respectively. We quantify the constants values by utilizing the Eqs. (\ref{5}), (\ref{12}) and (\ref{13}) that is
\begin{equation}\label{14a}
~~~~~a= \frac{(1+bR^{2})^{2}}{R}\sqrt{\frac{\frac{2M}{R}}{1-\frac{2M}{R}}}~,~~~~~~~~~~~~~~~~~~~~~~~b= \frac{4M-R}{R^{2}(9R-20M)}~,
\end{equation}
\begin{eqnarray}\label{14b}
A&=& \frac{aB}{2b(1+bR^{2})}+\sqrt{1-\frac{2M}{R}}~,~~~~~~~~~~~~~~~~B= (\frac{1}{2R})\sqrt{\frac{2M}{R}}~.
\end{eqnarray}
To determine the unknowns $a,~b,~A$ and $B$, we use the estimated radius and mass of stellar objects $LMC~X-4,$ $EXO~1785-248,$ $Cen~X-3$ and $4U~1820-30$ as shown in Tab. \ref{tab1}.
\begin{table}[ht]
\centering
\caption{Unknown constants compact stars $a$, $b$, $A$ and $B$}.
\begin{tabular}{|p{2cm}|p{2.5cm}| p{2.6cm}| p{2.5cm}| p{2.5cm}| p{2.5cm}|}
\hline
\hline
~
\begin{center}
\textbf{Star Model}
\end{center}
    & ~~~ \begin{center}
    $\textbf{LMC~X-4}$ \\
     $(S_{1})$
    \end{center} & ~~~  \begin{center}
    $\textbf{EX0~1785-248}$ \\
     $(S_{2})$
    \end{center} & ~~~ \begin{center}
    $\textbf{Cen~X-3}$ \\
     $(S_{3})$
    \end{center} & ~~~ \begin{center}
    $\textbf{4U~1820-30}$ \\
     $(S_{4})$
    \end{center}\\
\hline
~~~~ $M~(M_{\Theta})$ &~1.29 $\pm$ 0.05 \cite{Raw} &~1.30 $\pm$ 0.20 \cite{Oze} &~1.49 $\pm$ 0.08 \cite{Raw}   &~1.58 $\pm$ 0.06 \cite{TGuv}  \\
\hline
~~~~ $R~(km)$    &~~10.00 $\pm$ 0.11 &~~10.10 $\pm$ 0.44        &~~10.40 $\pm$ 0.15    &~~10.56 $\pm$ 0.10       \\
\hline
~~~~ $M/R$ &~~~~0.242338          &~~~~0.257507       &~~~~ 0.242338        &~~~~0.257507 \\
\hline
~~~~ $a~(km)$      &~~~~0.071746   &~~~~0.070870     &~~~~ 0.077520       &~~~~0.080474      \\
\hline
~~~~ $b~(km)$   &~~~-0.000453 &~~~-0.000447    &~~~~-0.000292    &~~~-0.000220 \\
\hline
~~~~ $A~(km)$     &~~~-1.776120    &~~~-1.753205   &~~~~-3.536640         &~~~-5.144660 \\
\hline
~~~~ $B~(km)$     &~~~~0.030929    &~~~~0.030588     &~~~~~0.031341       &~~~~0.031543\\
\hline
\end{tabular}
\label{tab1}
\end{table}
\FloatBarrier
Moreover, the mandatory conditions for well behaved compact structure are enlisted as
\begin{itemize}
  \item The graphs of $\rho,$ $p_{r}$ and $p_{t}$ must be positive, finite and maximum at center.
  \item The graphs of $\frac{d\rho}{dr},$ $\frac{dp_{r}}{dr}$ and $\frac{dp_{t}}{dr}$ must be negative.
  \item All the energy conditions must be satisfied.
  \item All the forces must fulfil the equilibrium stability condition.
  \item The ratios of the equation of state, $w_{r}$ and $w_{t}$, must be less than 1 i.e. $0<w_{r},~w_{t}<1.$
  \item Speed of sounds parameters ${v_{r}}^{2}$ and ${v_{t}}^{2}$ should lie within $[0,1]$.
  \item In case of anisotropic sphere, adiabatic index value must be larger then $4/3.$
\end{itemize}

\section{Physical features of $f(R,T)$ gravity model}
Next, we study the physical attributes of compact objects within the light of $f(R,T)$ gravity. To do so, we noticed the physical responses of some characteristics namely nature of energy density, pressure profile, energy bonds, surface redshift, equilibrium and stability conditions etc. Moreover, we take the specific values of parameters which are already presented in Tab. \ref{tab1}. Furthermore, to get the satisfactory outcomes of our chosen $f(R,T)$ theory model, we set the constant parameters, $\alpha$ and $\beta$ i.e. $\alpha=-7$ and $\beta=2.5$ for $LMC~X-4,$ $\alpha=-10$ and $\beta=2.6$ for $EXO~1785-248,$ $\alpha=-10$ and $\beta=1.8$ for $Cen~X-3$, $\alpha=-10$ and $\beta=1.6$ for $4U~1820-30$ and $\gamma=0.9$ is fix for all the stars.
\subsection{Energy Density and Pressure Profile}
\begin{figure}[h!]
\begin{tabular}{cccc}
\epsfig{file=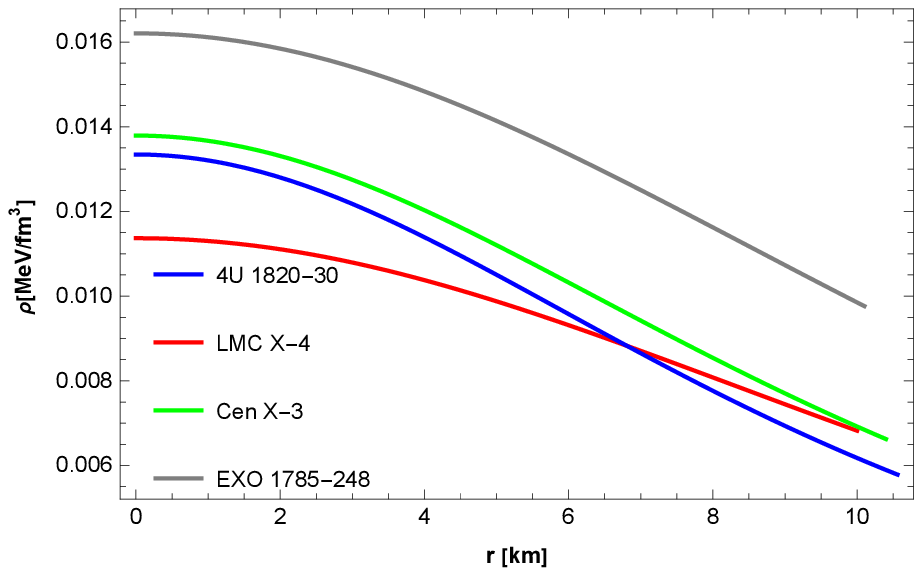,width=0.33\linewidth} &
\epsfig{file=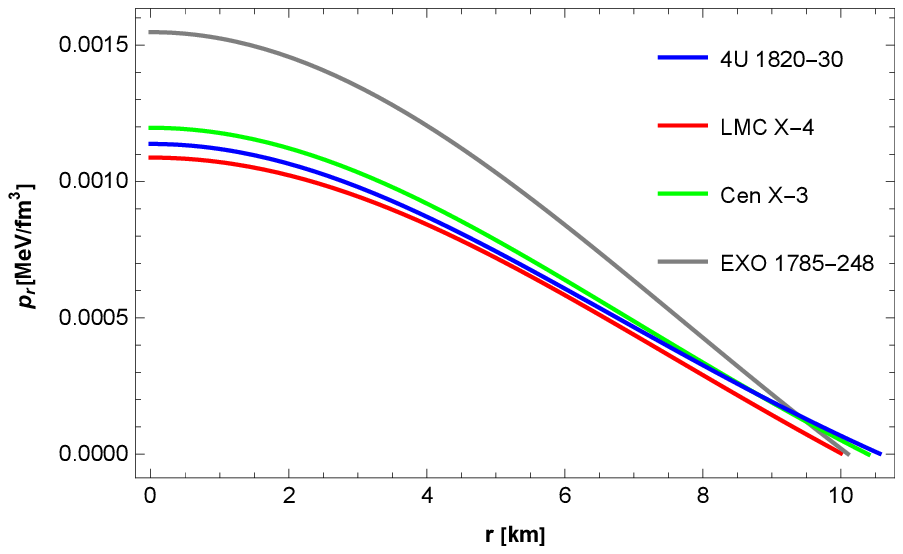,width=0.33\linewidth} &
\epsfig{file=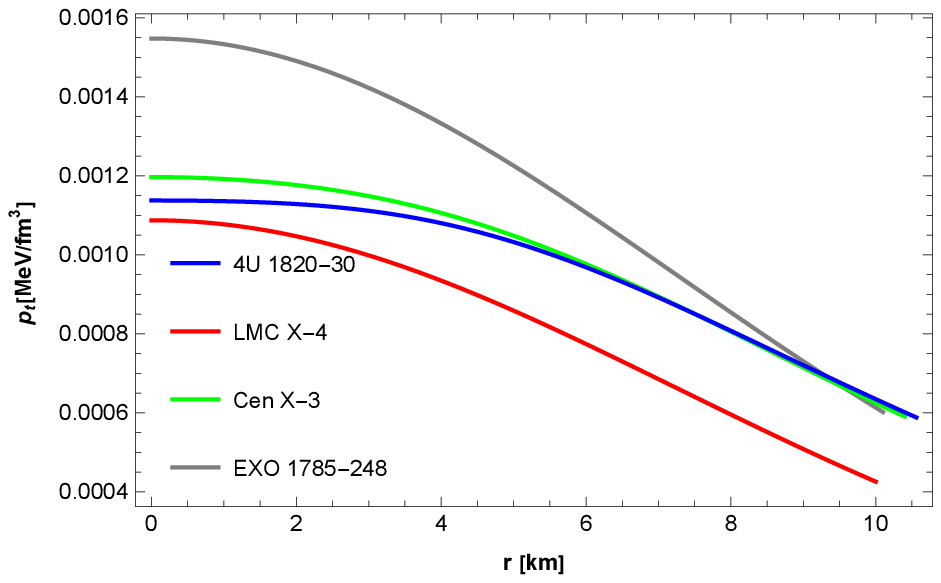,width=0.33\linewidth} &
\end{tabular}
\caption{{Graphical variations of $\rho$, $p_{r}$ and $p_{t}$}.}
\label{Fig:2}
\end{figure}

\begin{figure}[h!]
\begin{tabular}{cccc}
\epsfig{file=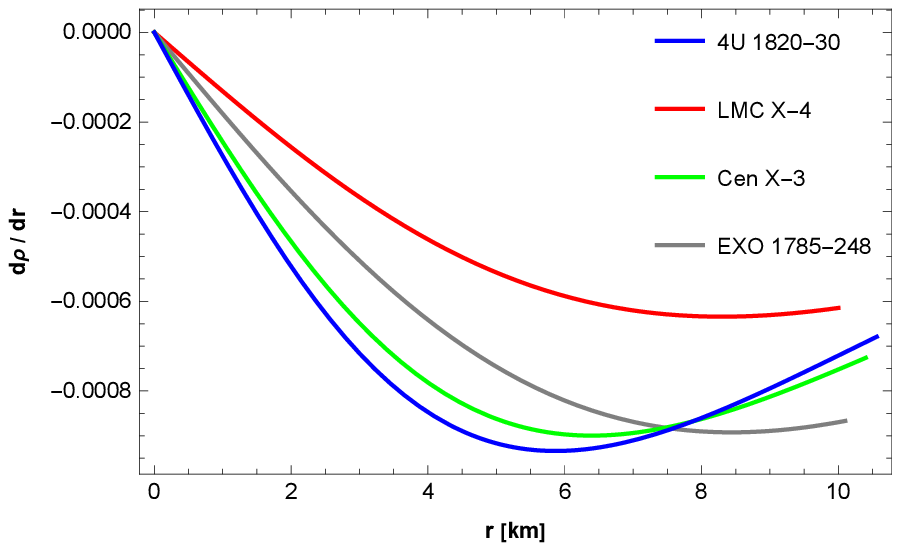,width=0.33\linewidth} &
\epsfig{file=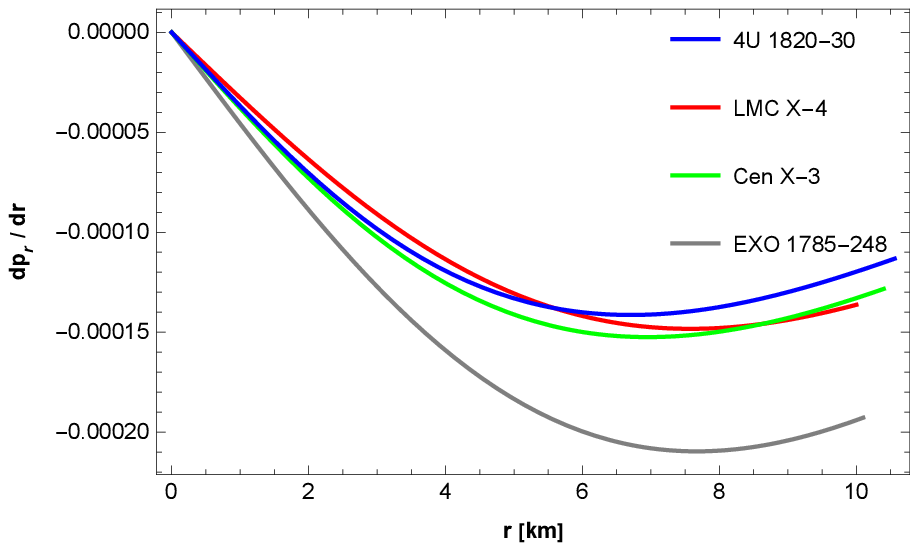,width=0.33\linewidth} &
\epsfig{file=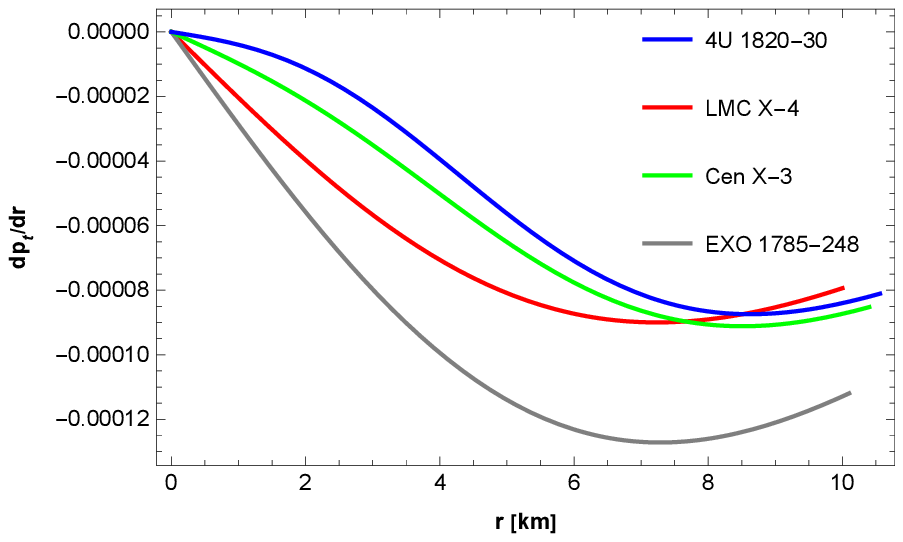,width=0.33\linewidth} &
\end{tabular}
\caption{{Evolution of $\frac{d\rho}{dr},$ $\frac{dp_{r}}{dr}$ and $\frac{dp_{t}}{dr}$}.}
\label{Fig:3}
\end{figure}

Firstly, we provide the graphical illustrations of energy density, radial pressure and tangential pressure as given in Fig. $\ref{Fig:2}$. It’s been observed that these plots are positive and maximum at the core of the compact objects. This graphical response of $\rho$, $p_{r}$ and $p_{t}$ ensures that our system is stabilized.
\FloatBarrier
Moreover, the graphical illustrations of gradients of energy density and pressure components are also presented in our analysis. The satisfactory outcomes of these plots, given in Fig. $\ref{Fig:3},$ reflects the consistent nature of our system.
\FloatBarrier
\subsection{Anisotropy}
\begin{figure}[h!]
\begin{tabular}{cccc}
\epsfig{file=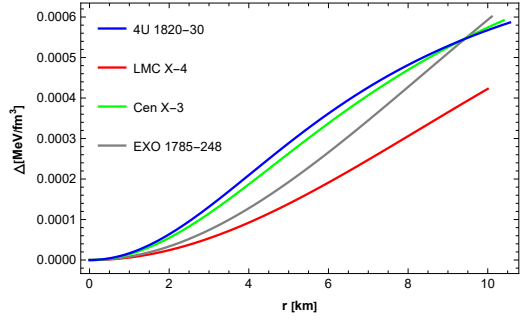,width=0.38\linewidth} &
\end{tabular}
\caption{{Evolution of anisotropy}.}
\label{Fig:4}
\end{figure}

In this section, we describe the graphical evolution of anisotropic factor \cite{BH} for our stellar system. It has been observed that for the spherically symmetric matter distribution the anisotropic factor exhibits positive nature throughout the star and is defined as $\Delta=p_t-p_r$. One can notice from Fig. $\ref{Fig:4}$ that the graphical response of anisotropy reveals positive and repulsive nature \cite{MK}.
\FloatBarrier
\subsection{Energy Conditions}
\begin{figure}[h!]
\begin{tabular}{cccc}
\epsfig{file=Density.eps,width=0.33\linewidth} &
\epsfig{file=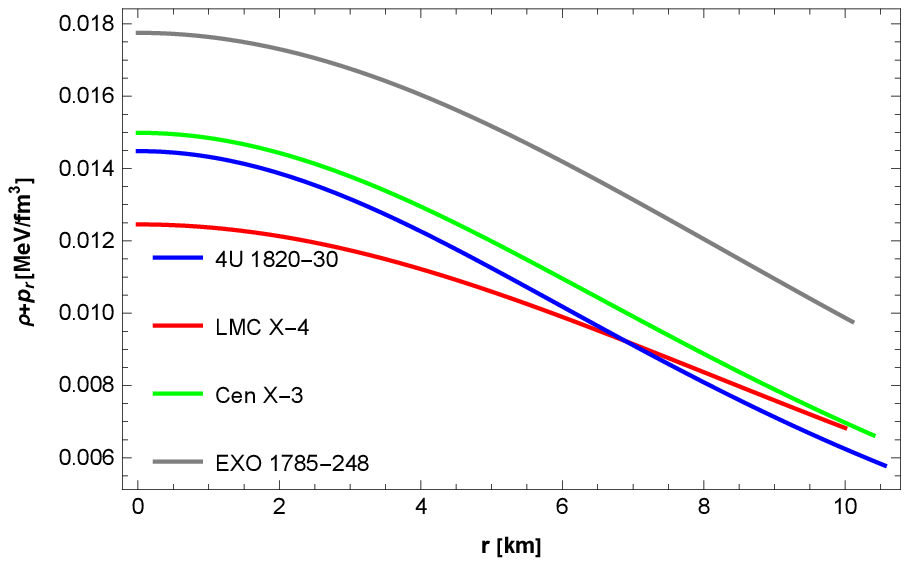,width=0.33\linewidth} &
\epsfig{file=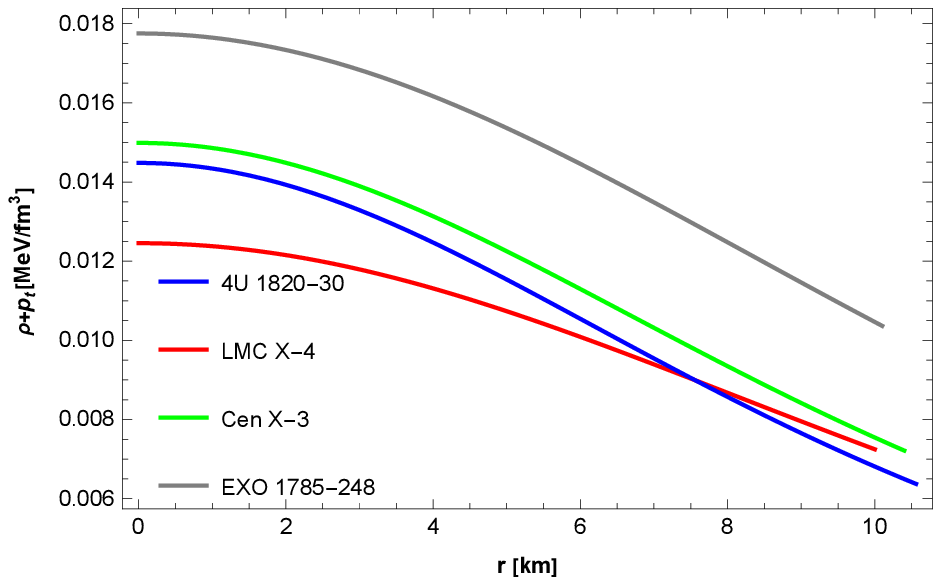,width=0.33\linewidth} &\\
\epsfig{file=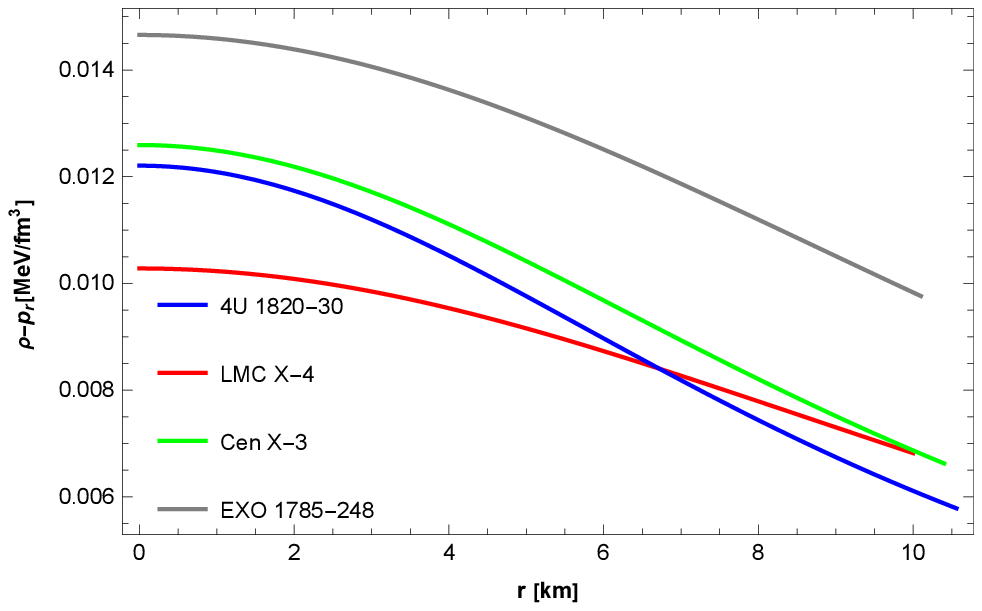,width=0.33\linewidth} &
\epsfig{file=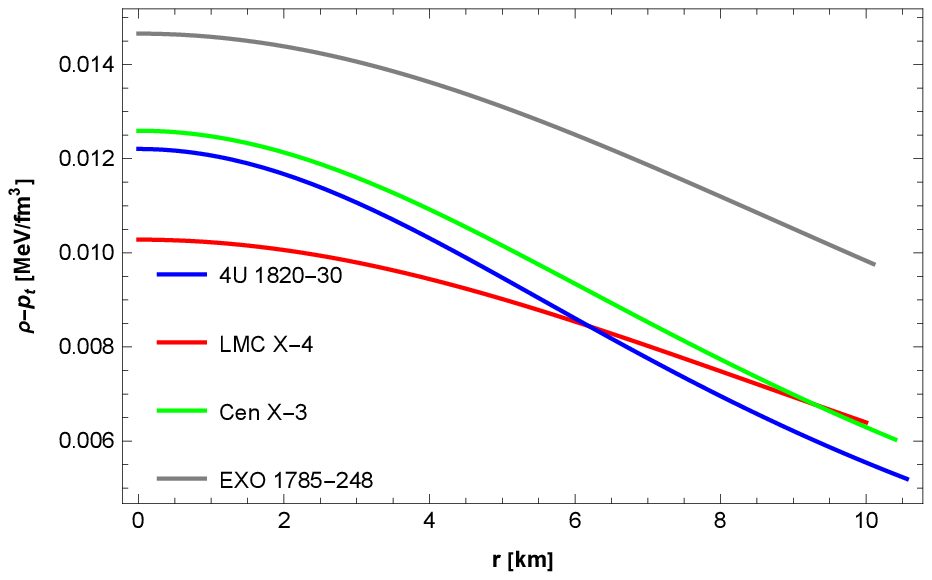,width=0.33\linewidth} &
\epsfig{file=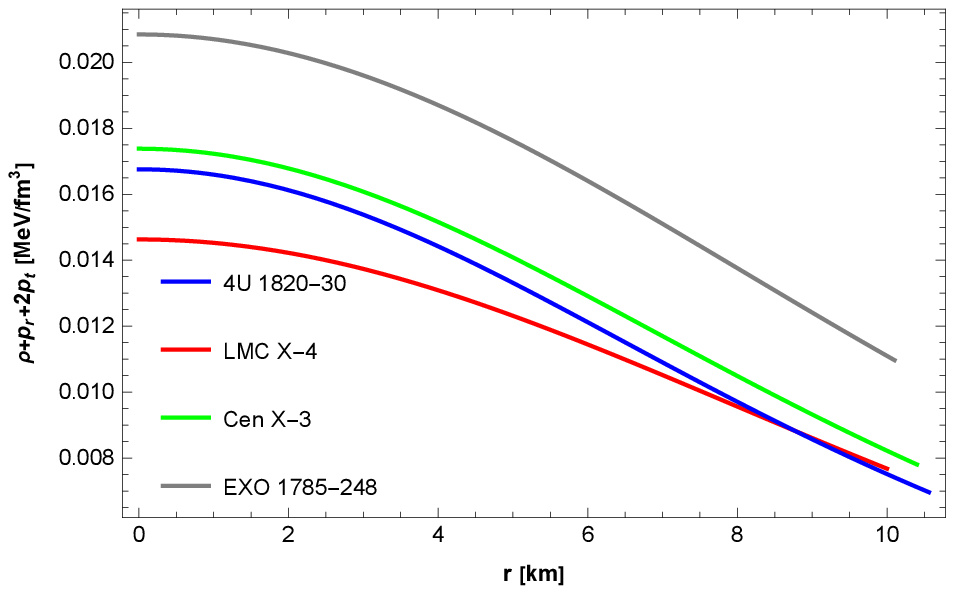,width=0.33\linewidth} &
\end{tabular}
\caption{{Graphs of energy conditions}.}
\label{Fig:5}
\end{figure}

Next, we study the graphical illustration of different energy attributes.
These attributes are popularly classified into the four types and are defined as $NEC$ as null energy condition, $WEC$ as weak energy condition, $SEC$ as strong energy condition and $DEC$ as dominant energy condition.
\begin{equation*}
NEC:  \rho+p_{r}\geq 0,\rho+p_{t},~~~~~~~~WEC:\rho \geq,\rho+p_{r} \geq 0,\rho+p_{t} \geq 0,
\end{equation*}
\begin{equation*}
SEC: \rho+p_{r}\geq 0,\rho+p_{t}\geq 0, \rho+p_{r}+2p_{t}\geq 0,~~~~~~~~DEC:\rho\geq,\rho\pm p_{r} \geq 0,\rho\pm p_{t} \geq 0.
\end{equation*}
From Fig. $\ref{Fig:5}$, one can clearly notice that these energy attributes show the well fitted behavior for our proposed model.
\FloatBarrier
\subsection{Equilibrium Condition}
\begin{figure}[h!]
\begin{tabular}{cccc}
\epsfig{file=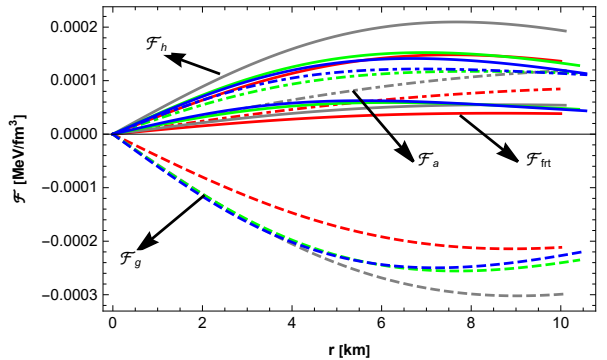,width=0.45\linewidth} &
\end{tabular}
\caption{{Behavior of $\mathcal{F}_{h}$, $\mathcal{F}_{g}$, $\mathcal{F}_{a}$}, and $\mathcal{F}_{frt}$.}
\label{Fig:6}
\end{figure}

Moreover, the equilibrium stability is considered as the key feature in the analysis of compact stars. Here, we study the equilibrium constraints for our stellar system in the light of anisotropic sphere. In case of $f(R,T)$ gravity, the famous $TOV$ \cite{Tol,opp1} equation takes the succeeding modified form
\begin{equation}\label{15}
 \frac{\nu^{'}}{r}(\rho+p_{r})+\frac{dp_{r}}{dr}-\frac{2}{r}(p_{t}-p_{r})+\frac{2\gamma}{3(8\pi-\gamma)}\frac{d}{dr}(3\rho-p_{r}-2 p_{t})=0.
\end{equation}
Here, $\mathcal{F}_{g}=-\frac{\nu^{'}}{r}(\rho+p_{r})$ indicates gravitational force, $\mathcal{F}_{h}=-\frac{dp_{r}}{dr}$ represents the hydrostatic force, $ \mathcal{F}_{a}=\frac{2}{r}\Delta$ denotes anisotropic force and $\mathcal{F}_{frt}=-\frac{2\gamma}{3(8\pi-\gamma)}\frac{d}{dr}(3\rho-p_{r}-2 p_{t})$ is the additional force due to the modification in $f(R,T)$ gravity. For the stellar structure to be equilibrium, the aggregation of these forces must be zero. Thus,
$$\mathcal{F}_{g}+\mathcal{F}_{h}+\mathcal{F}_{a}+\mathcal{F}_{frt}=0.$$
Fig. $\ref{Fig:6}$ shows that our system satisfies the necessary equilibrium condition, which highlights that our chosen model is well fitted.
\FloatBarrier
\subsection{Equation of State}
\begin{figure}[h!]
\begin{tabular}{cccc}
\epsfig{file=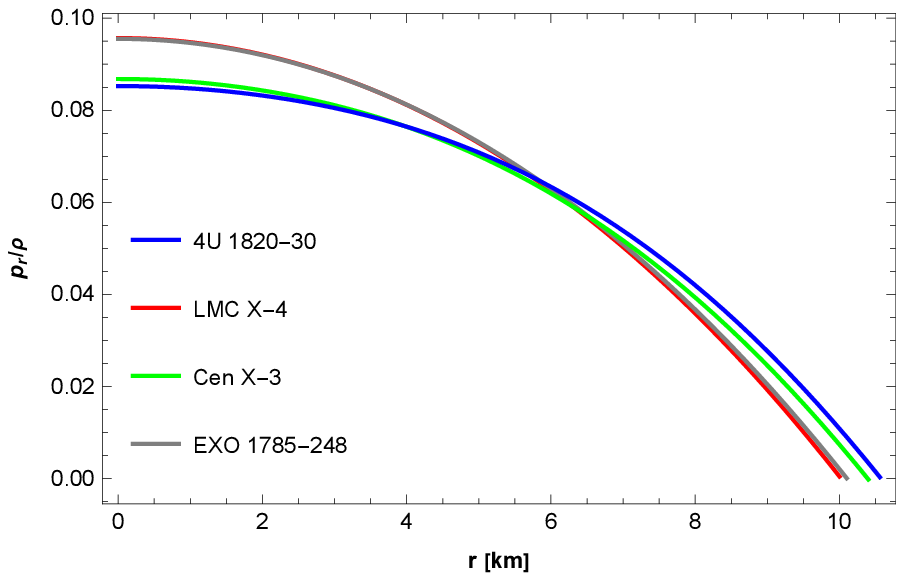,width=0.38\linewidth} &
\epsfig{file=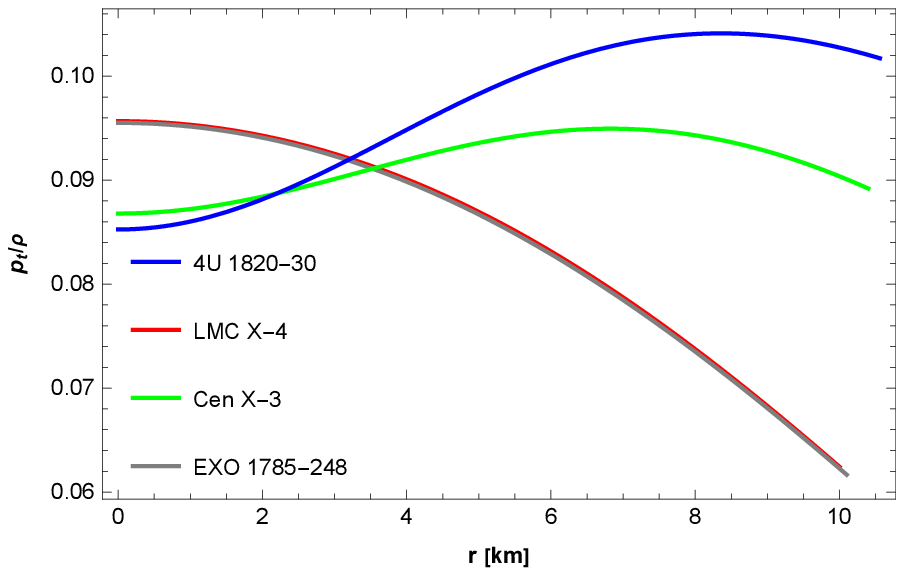,width=0.38\linewidth} &
\end{tabular}
\caption{{Evolution of $w_{r}$ and $w_{t}$}.}
\label{Fig:7}
\end{figure}

Further, we examine the graphical response of two ratios of equation of state $(EoS)$ which are  denoted $w_{r}$ and $w_{t}$ and are defined by two fractions i.e.
\begin{equation}\label{26}
   w_{r} = \frac{p_{r}}{\rho},~~~~~~~~w_{t} =  \frac{p_{t}}{\rho}.
\end{equation}
For the well fitted behavior the graphical illustration of the compact star must lie in the range of $0$ and $1$ i.e. $0<w_{r},~w_{t}<1.$ It can be easily observed from Fig. $\ref{Fig:7}$ that the satisfactory behaviour of $EoS$ confirms the consistency of our $f(R,T)$ gravity model.
\FloatBarrier
\subsection{Causality Condition}
\begin{figure}[h!]
\begin{tabular}{cccc}
\epsfig{file=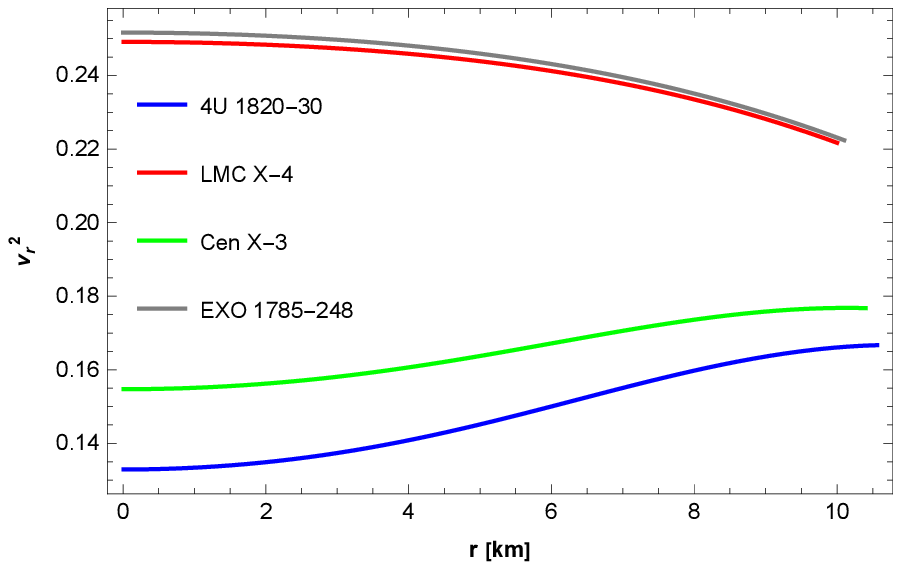,width=0.33\linewidth} &
\epsfig{file=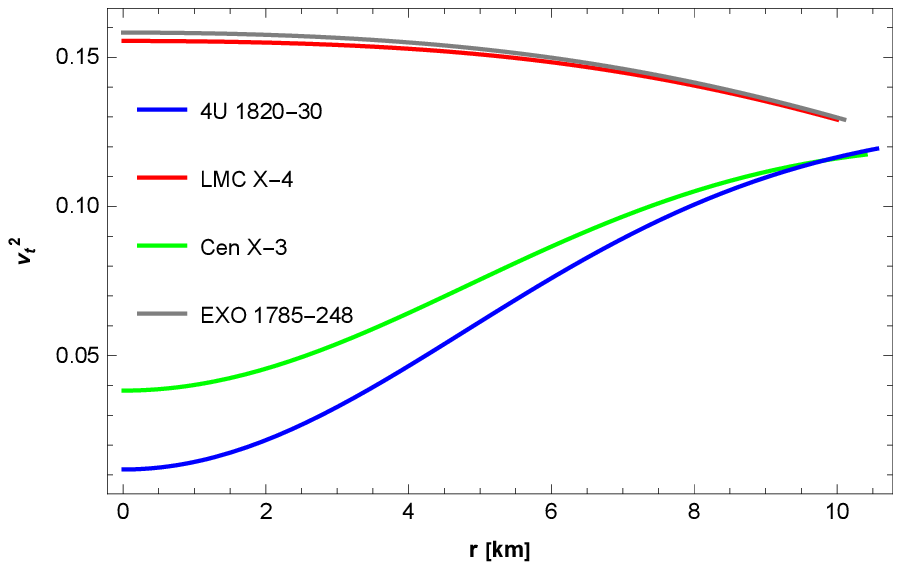,width=0.33\linewidth} &
\epsfig{file=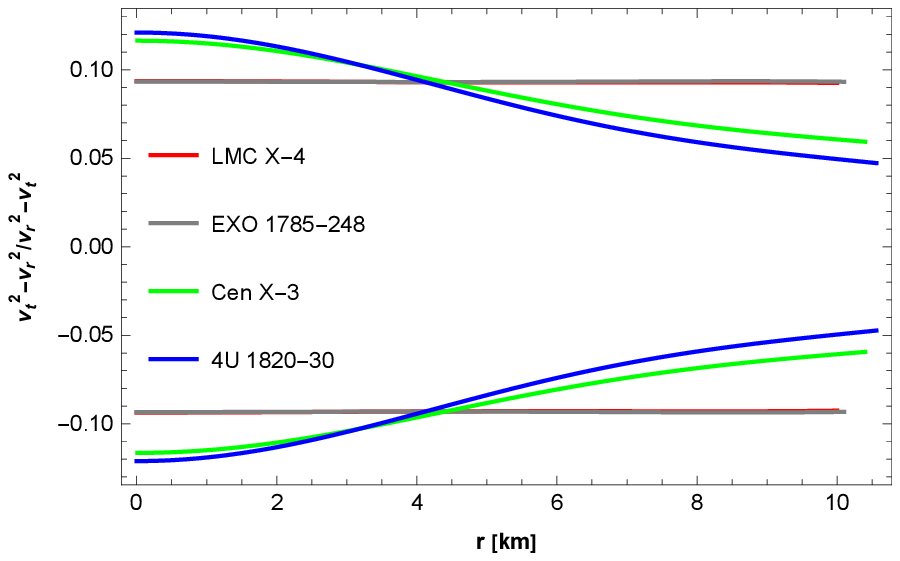,width=0.33\linewidth} &
\end{tabular}
\caption{{Behavior of ${v_{r}}^{2},$ ${v_{t}}^{2}$ and $|{v_{r}}^{2}-{v_{t}}^{2}|$ for $f(R,T)$ gravity Model}.}
\label{Fig:8}
\end{figure}

In the physical analysis of the compact star the stability condition plays a major role. Here, we apply the Herrera cracking concept \cite{Her}, to probe the stability of chosen model. Generally, the speed of sound parameters parted into radial component ${v_{r}}^{2}$ and transversal component ${v_{t}}^{2}$ and are describe as
\begin{equation}\label{27}
  {v_{r}}^{2} = \frac{dp_{r}}{d\rho},~~~~~~~~{v_{t}}^{2} = \frac{dp_{t}}{d\rho}.
\end{equation}
According to Herrera, the range of sound speed ${v_{r}}^{2}$ and ${v_{t}}^{2}$ must varies in between $0$ and $1.$ Abreu et al. \cite{Abr} introduced another important concept to investigate the possibly stable/unstable structures of the celestial objects. The variations in speed of sound affect potentially stable/unstable areas within matter distributions. The region of an anisotropic matter distribution where $0\leq{v_{t}}^{2}-{v_{r}}^{2}\leq1.$ It indicates that the horizon in which the transverse speed parameter is less than the radial speed sound parameter and is acknowledged as stable otherwise not. From Fig. $\ref{Fig:8}$ one can notice that all the mentioned stability conditions are well satisfied.
\FloatBarrier
\subsection{Analysis of Redshift, Mass Function and  Compactness Factor}
\begin{figure}[h!]
\begin{tabular}{cccc}
\epsfig{file=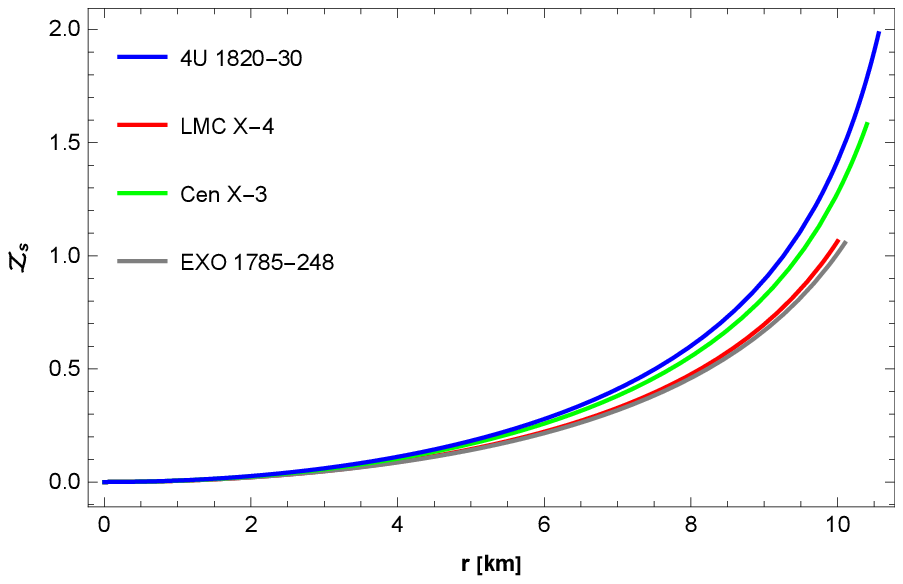,width=0.33\linewidth} &
\epsfig{file=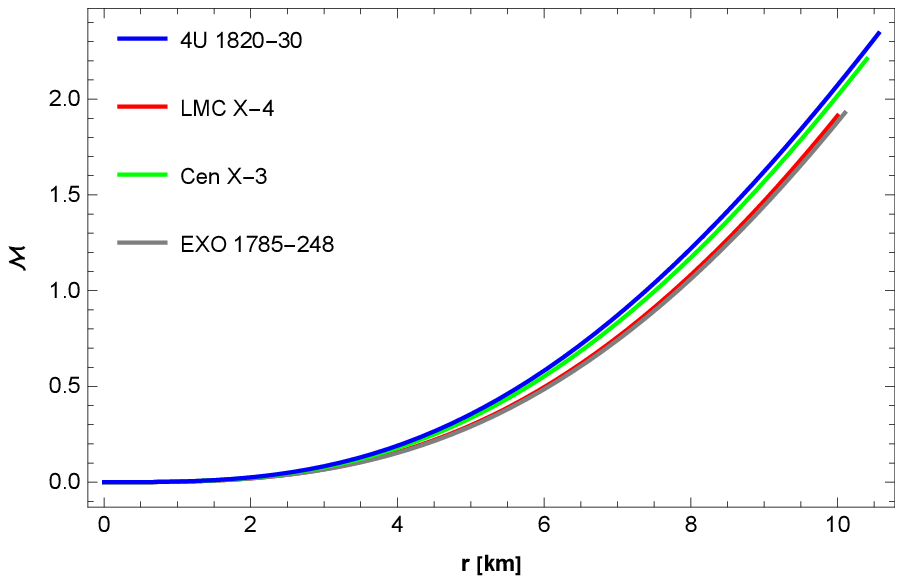,width=0.33\linewidth} &
\epsfig{file=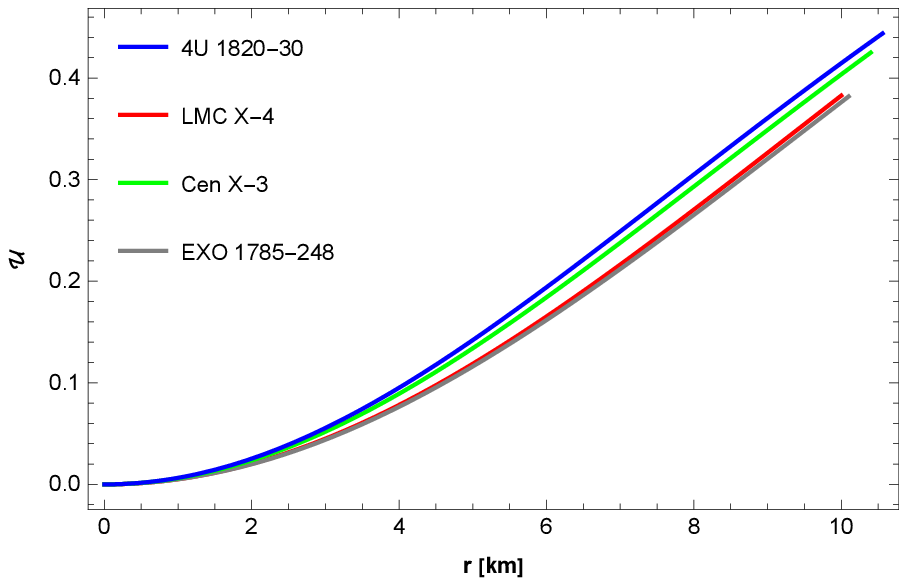,width=0.33\linewidth} &
\end{tabular}
\caption{{Behavior of surface redshift (left panel), mass function (middle panel) and compactness factor (right panel)}.}
\label{Fig:9}
\end{figure}

The surface redshift \cite{Boh} of the compact star is described as
\begin{equation}
  \text{$\mathcal{Z}_s$} = \frac{1}{\sqrt{1-2\nu}}-1.
\end{equation}
Further, Buchdahl \cite{BB} examined the idea of the highest acceptable mass-radius ratio in the context of an isotropic fluid, i.e., $2$ $\frac{M}{R}< \frac{8}{9}$. Then Mak and Harko \cite{MAK} elaborated on the Buchdahl concept and claimed that this mass-radius ratio applies to both isotropic and anisotropic fluid. The expression mass function \cite{BSM} is obtained by employing the metric potentials $g_{rr}^{-}$ = $g_{rr}^{+}$
\begin{equation}
 \mathcal{M}(r)=\frac{a^{2}r^{3}}{2((1+br^{2})^{4}+a^{2}r^{2})}.
\end{equation}
The nature of mass function presents that mass is regular at core, i.e. $\mathcal{M}(r) \rightarrow 0.$ Moreover, the mathematical form of compactness parameter $\mathcal{U}(r)$ \cite{MAK} is defined by
\begin{equation}
\mathcal{U}(r)=\frac{2\mathcal{M}(r)}{r}=\frac{a^{2}r^{2}}{[(1+br^{2})^{2}+a^{2}r^{2}]}.
\end{equation}
Fig. $\ref{Fig:9}$ demonstrates the physical illustrations of surface redshift, mass function and compactness factor. It can be noticed that the graphical response of all these three parameters exhibit monotonically increasing behavior for our chosen $f(R,T)$ gravity model.
\FloatBarrier
\subsection{Adiabatic Index}
\begin{figure}[h!]
\begin{tabular}{cccc}
\epsfig{file=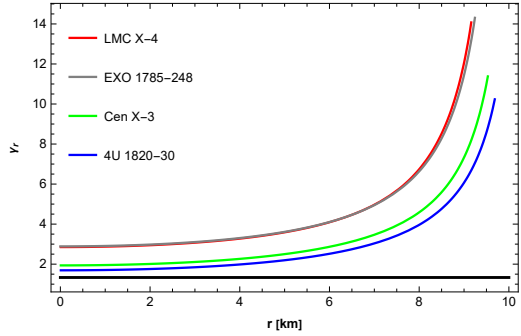,width=0.38\linewidth}
\end{tabular}
\caption{{Behavior of  $\gamma_{r}$}.}
\label{Fig:10}
\end{figure}

Further, in this portion, we explore the stiffness of $EoS$ by studying the adiabatic index. For a stabilized configuration, adiabatic index value must be larger then $4/3$. The ratio of adiabatic index is represented as
\begin{equation}\label{26}
\gamma_{r}=\frac{\rho+p_{r}}{p_{r}}(\frac{dp_{r}}{d\rho})=\frac{\rho+p_{r}}{p_{r}}{v_{r}}^{2}.
\end{equation}
It has been observed from Fig. $\ref{Fig:10}$ that our chosen $f(R,T)$ gravity model reveals the stabilized behaviour of the compact star.
\FloatBarrier
\section{Conclusion}
In order to explore the complicated nature of the compact star, different modified theories is considered as key aspect in astrophysics. In these modified theories, $f(R,T)$ gravity gains most attention among cosmologists because of its special combination of curvature and matter. Moreover, $f(R,T)$ presented some remarkable results in the field of cosmology, thermodynamic and astrophysics of compact objects \cite{Jam1,Jam2,Shab1,Shab2}. Many fruitful outcomes of $f(R,T)$ gravity had already been explored in different frames of cosmic dynamics. In our present study, we examined the exact solution of the $f(R,T)$ field equation to understand the behavior of stellar stars along with anisotropic fluid source. To achieve our goal, we use the remarkable approach known as Karmarkar condition \cite{KR}, which reduces the solution-generating approach into a single metric potential by relating $g_{tt}$ with $g_{rr}$. Moreover, we consider the specific expression for the potential $e^{\lambda}$ \cite{BSM}, i.e. $e^{\lambda}=1+ \frac{a^{2}r^{2}}{(1+br^{2})^{4}}$ and obtained the value of other metric potential by utilizing the Karmarkar condition, given as  $e^{\nu}={\Big[A-\frac{aB}{2b(1+br^{2})}\Big]}^{2}$. Here $A$, $B$, $a$ and $b$ are any unknowns. Furthermore, we apply junction conditions to calculate the values of unknowns by comparing the interior sphere (\ref{5}) with the Schwarzschild exterior geometry (\ref{12}).
To validate the stability of our obtained solutions, we examine the graphical responses of four different compact objects, namely $LMC~X-4$, $EXO~1785-248$, $Cen~X-3$ and $4U~1820-30$. The aim of our study is to create a family of embedded class-I solutions of the $f(R,T)$ theory in the light of anisotropic fluid source. The noteworthy results are enlisted beneath.
\begin{itemize}
    \item Graphical nature of both metric potentials shown in Fig. $\ref{Fig:1}$ indicates that $g_{tt}$ and $g_{rr}$ are free of singularity, positive and satisfy the constraints, i.e. $e^{\nu(r=0)}={\Big[A-\frac{aB}{2b}\Big]}^{2}$ and $e^{\lambda(r=0)}=1$. The behavior of both the metric potential highlights satisfactory outcomes, as their plots increase monotonically and reach the maximum values at the boundary.
    \item Fig. $\ref{Fig:2}$ reveals the consistent nature of density and pressure components for the model under discussion. The graphical response of these plots reaches the highest value in the center and transmit decreasing behavior towards the boundary.
    \item The gradient of pressure components and energy density can be observed from Fig. $\ref{Fig:3}$. These plots exhibit negative illustrations which confirm that our obtained results are stable.
    \item The energy bonds for chosen $f(R,T)$ model are presented in Fig. $\ref{Fig:5}$. One can note that our chosen model fulfills all these necessary constraints.
    \item The graphical illustration of the combined forces i.e. $\mathcal{F}_{a}$, $\mathcal{F}_{g}$ $\mathcal{F}_{h}$ and $\mathcal{F}_{frt}$, presented in Fig. $\ref{Fig:6}$, reconfirm the validity of our stellar system.
    \item The graphical response of anisotropy, shown in Fig. $\ref{Fig:4}$, exhibits the repulsive and well behaved nature of our system.
    \item It is noted from Fig. $\ref{Fig:7}$ that the nature of both the ratios of $EoS$ is stable.
    \item Moreover, the graphical responses of both the speed of sound parameters, $v_{r}$ and $v_{t},$ have also been studied in our analysis. Since for a well fitted stellar structure, the plotting range of both the speed of sound parameters must lie within the value of $0$ and $1$. It can clearly be noted from Fig. $\ref{Fig:8}$ that both the plots of $v_{r}$ and $v_{t}$ successfully satisfied this condition.
    \item One can easily observe the monotonically increasing nature of surface redshift, mass function and compactness factor from Fig. $\ref{Fig:9}.$
    \item The satisfactory behavior of adiabatic index can be noticed from Fig. $\ref{Fig:10}$.
\end{itemize}
Thus, our analysis showed a stable nature for all the necessary constraints and exhibit well fitted behavior in the light of anisotropic fluid source. Moreover, it is valuable to highlight that our proposed results are almost identical to the outcomes explored by Naz et al. \cite{Naz} in $f(R)$ gravity background.


\begin{thebibliography}{70}
\bibitem{Sch}  K. Schwarzschild, Sit. Dts. Aka. Wis. Math. Phys. Ber. \textbf{24}, 424 (1916).
\bibitem{Tol} R. C. Tolman, Phys. Rev. \textbf{55}, 364 (1939).
\bibitem{opp1} J. R. Oppenheimer, G. M. Volkoff, Phys. Rev. \textbf{55}, 374 (1939).
\bibitem{Baad} W. Baade, F. Zwicky, Phys. Rev. \textbf{46}, 76 (1934).
\bibitem{Bow} R. L. Bowers, E. P. T. Liang, Astrophy. J. \textbf{188}, 657 (1974).
\bibitem{Rud} R. Ruderman, Ann. Rev. Ast. Astrophys. \textbf{10}, 427 (1972).
\bibitem{Saw} R. F. Sawyer, Phys. Rev. Lett. \textbf{29}, 6, 382–385 (1972)..
\bibitem{Put} A. Putney, APJL \textbf{451}, 67 (1995).
\bibitem{Rei} D. Reimers et al., Astron. Astrophys. \textbf{311}, 572 (1996).
\bibitem{Mar} A. P. Martinez, R. G. Felipe, D. M. Paret, Int. J. Mod. Phys. D \textbf{19}, 1511 (2010).
\bibitem{Sham1} M. F. Shamir, M. Ahmad, Mod. Phys. Lett. A \textbf{32}, 1750086 (2017).
\bibitem{Sham2} M. F. Shamir, M. Ahmad, Eur. Phys. J. C \textbf{77}, 55 (2017).
\bibitem{BLi2} B. Li, T. P. Sotiriou,  J. B. Barrow, Phys. Rev. D \textbf{83}, 064035 (2011).
\bibitem{Har} T. Harko, F. S. N. Lobo, S. Nojiri, S. D. Odintsov, Phys. Rev. D \textbf{84}, 024020 (2011).
\bibitem{Cap2} S. Capozziello, M. D. Laurentis, S. D. Odintsov, A. Stabile, Phys. Rev. D \textbf{83}, 064004 (2011).
\bibitem{Lov} D. Lovelock, J. Math. Phys. \textbf{12}, 498 (1971).
\bibitem{Har2}  T. Harko, Mon. Not. R. Astron. Soc. \textbf{413}, 3095 (2011).
\bibitem{Ikr}  M. Sharif, A. Ikram, J. Exp. Theor. Phys. \textbf{123}, 40 (2016).
\bibitem{ad5} M. F. Shamir and A. Malik, Chin. J. Phys. \textbf{69}, 312 (2021).
\bibitem{ad6} M. F. Shamir and A. Malik, Commun. Theor. Phys. 71, (2019).
\bibitem{ad7} A. Malik, F. Mofarreh, A. Zia and A. Ali, Chinese Phys. C, \textbf{46}, 095104  (2022).
\bibitem{ad9} A. Malik, Eur. Phys. J. Plus \textbf{136}, 1146 (2021).
\bibitem{Zoy} M. F. Shamir, Z. Asghar, A. Malik, Fortschr. Phys. 2200134 (2022).
\bibitem{Buch} H. A. Buchdahl, Mon. Not. R. Ast. Soc. \textbf{150}, 1 (1970).
\bibitem{Adh} K. S. Adhav, Astrophys. Space Sci. \textbf{339}, 365-369 (2012).
\bibitem{Nai} R. L. Naidu, D. R. K. Reddy, T. Ramprasad, K. V. Ramana, Astrophys. Space Sci. \textbf{348}, 247 (2013).
\bibitem{Sha1}  M. Sharif, M. Zubair, Astrophys. Space Sci. \textbf{349}, 529 (2014).
\bibitem{Zub} M. Sharif, M. Zubair, JCAP \textbf{03}, 028 (2012).
\bibitem{Pre} J. M. Z. Pretel, S. E. Joras, R. R. R. Reis, J. D. V. Arbanil, JCAP \textbf{08}, 055 (2021).
\bibitem{WGZA} S. Waheed, G. Mustafa, M. Zubair, A. Ashraf, Sym. \textbf{12}, 6, 962 (2020).
\bibitem{Schl} Schlai, Ann. di Mat. \textbf{5}, 170 (1871).
\bibitem{Nas} J. Nash, Ann. Math. \textbf{63}, 20 (1956).
\bibitem{Gunt} M. Gunther, Math. Nachr. \textbf{144}, 165 (1989).
\bibitem{Gup} Y. K. Gupta, J. R. Sharma, Gen. Rel. Grav. \textbf{28}, 1447 (1996).
\bibitem{KR} K. R. Karmarkar, Proc. Ind. Aca. Sci. A \textbf{27}, 56 (1948).
\bibitem{Naz} T. Naz, A. Usman, M. F. Shamir, Ann. Phys. \textbf{429}, 168491 (2021).
\bibitem{Cog} G. Cognola et al., Phys. Rev. D \textbf{77}, 046009 (2008).
\bibitem{BSM} P. Bhar, K. N. Singh, T. Manna, Int. J. Mod. Phys. D \textbf{26}, 1750090 (2017).
\bibitem{Barri} O. J. Barrientos, G. F. Rubilar, Phys. Rev. D \textbf{90}, 028501 (2014).
\bibitem{Fara} V. Faraoni, Phys. Rev. D \textbf{81}, 044002 (2010).
\bibitem{AVC} A. V. Astashenok, S. Capozziello, S. D. Odintsov, J. Cos. Ast. Phys. \textbf{12}, 040 (2013).
\bibitem{Coon} A. Cooney, S. D. Deo, D. Psaltis, Phys. Rev. D \textbf{82}, 064033 (2010).
\bibitem{Gang} A. Ganguly, R. Gannouji, R. Goswami, S. Ray, Phys. Rev. D \textbf{89}, 064019 (2014).
\bibitem{Mome} D. Momeni, R. Myrzakulov, Int. J. Geo. Met. Mod. Phys. \textbf{12}, 1550014 (2015).
\bibitem{Raw} M. L. Rawls, et al., APJ, \textbf{25}, 730 (2011).
\bibitem{Oze} F. Ozel, T. Guver, T. Psaltis, APJ \textbf{693}, 1775 (2009).
\bibitem{TGuv} T. Guver et al., Astrophys J. \textbf{719}, 1807 (2010).
\bibitem{BH} C. G. Bohmer, T. Harko, Cla. Qua. Grav. \textbf{23}, 6479 (2006).
\bibitem{MK} M. K. Gokhroo, A. L. Mehra, Gen. Rel. Grav. \textbf{26}, 75 (1994).
\bibitem{BB} H. A. Buchdahl, Phys. Rev. \textbf{116}, 1027 (1959).
\bibitem{MAK} M. K. Mak, T. Harko, Proc. R. Soc. Lond. A \textbf{459}, 393 (2003).
\bibitem{Her} L. Herrera, Phys. Lett. A  \textbf{165}, 206 (1992).
\bibitem{Abr} H. Abreu, H. Hernandez, and L. A. Nunez,Class Quantum Grav. \textbf{24}, 4631 (2007).
\bibitem{And} H. Andreasson, Com. Math. Phys. \textbf{288}, 715 (2009).
\bibitem{Boh} C. G. Bohmer, T. Harko, Class. Quant. Grav. \textbf{23}, 6479 (2006).
\bibitem{Mak} M. K. Mak, T. Harko, Proc. R. Soc. Lon. A \textbf{459}, 393 (2003).
\bibitem{Jam1} M. Jamil, D. Momeni, R. Myrzakulov, Eur. Phys. J. C. \textbf{72}, 1959 (2012).
\bibitem{Jam2} M. Jamil, D. Momeni, R. Myrzakulov, Chin. Phys. Lett. \textbf{29}, 109801 (2012).
\bibitem{Shab1}  H. Shabani, M. Farhoudi, Phys. Rev. D \textbf{88}, 044048 (2013).
\bibitem{Shab2} H. Shabani, M. Farhoudi, Phys. Rev. D \textbf{90}, 44031 (2014).

\end{thebibliography}
\end{document}